\documentclass[12pt]{JHEP}

\font\blackboard=msbm10 at 12pt
\font\blackboards=msbm7
\font\blackboardss=msbm5
\newfam\black
\textfont\black=\blackboard
\scriptfont\black=\blackboards
\scriptscriptfont\black=\blackboardss
\def\bb#1{{\fam\black\relax#1}}

\newcommand{\NP}{{\em Nucl.\ Phys.\ }}

\newcommand{\PR}{{\em Phys.\ Rev.\ }}

\newcommand{\ba}{\begin{array}}
\newcommand{\ea}{\end{array}}
\newcommand{\be}{\begin{equation}}
\newcommand{\ee}{\end{equation}}
\newcommand{\bea}{\begin{eqnarray}}
\newcommand{\eea}{\end{eqnarray}}
\newcommand{\beas}{\begin{eqnarray*}}
\newcommand{\eeas}{\end{eqnarray*}}

  \def\CD{{\cal D}}

\def\darr#1{\raise1.5ex\hbox{$\leftrightarrow$}\mkern-16.5mu #1}

\def\half{{\textstyle{1\over2}}} 
\def\roughly#1{\raise.3ex\hbox{$#1$\kern-.75em\lower1ex\hbox{$\sim$}}}

\def\IB{\relax\hbox{$\inbar\kern-.3em{\rm B}$}}
\def\IC{\relax\hbox{$\inbar\kern-.3em{\rm C}$}}
\def\ID{\relax\hbox{$\inbar\kern-.3em{\rm D}$}}
\def\IE{\relax\hbox{$\inbar\kern-.3em{\rm E}$}}
\def\IF{\relax\hbox{$\inbar\kern-.3em{\rm F}$}}
\def\IG{\relax\hbox{$\inbar\kern-.3em{\rm G}$}}
\def\IGa{\relax\hbox{${\rm I}\kern-.18em\Gamma$}}
\def\IH{\relax{\rm I\kern-.18em H}}
\def\IK{\relax{\rm I\kern-.18em K}}
\def\IL{\relax{\rm I\kern-.18em L}}
\def\IP{\relax{\rm I\kern-.18em P}}
\def\IR{\relax{\rm I\kern-.18em R}}
\def\IZ{\relax\ifmmode\mathchoice{
\hbox{\cmss Z\kern-.4em Z}}{\hbox{\cmss Z\kern-.4em Z}}
{\lower.9pt\hbox{\cmsss Z\kern-.4em Z}}
{\lower1.2pt\hbox{\cmsss Z\kern-.4em Z}}
\else{\cmss Z\kern-.4em Z}\fi}
\def\II{\relax{\rm I\kern-.18em I}}

\def\ee#1{{\rm erf}\left(#1\right)}

\def\CD{{\cal D}}

\def\p{\partial}

\def\Tr{{\rm Tr}}

\def\inbar{\,\vrule height1.5ex width.4pt depth0pt}
\def\d{\delta}
\font\cmss=cmss10 \font\cmsss=cmss10 at 7pt

\def\d{{\delta}}

\def\s{{\sigma}}

\def\lref{\begingroup\obeylines\lr@f}
\def\lr@f#1#2{\gdef#1{\ref#1{#2}}\endgroup\unskip}
\def\s{\sigma}
\def\ad{a^\dagger}
\def\an{a_n}
\def\and{{a^\dagger_n}}
\def\vac{{|\Omega\rangle}}
\def\P{\Psi}

\def\math@note#1{\gdef\@eqnlabel{LAB: #1}}
\title{Split string field theory I}
\author{David J. Gross\\
{Institute for Theoretical Physics}\\
{University of California, Santa Barbara}\\
{Santa Barbara, CA 93106, U.S.A.}\\
{\tt gross@itp.ucsb.edu}}
\author{Washington Taylor\footnote{Current
address:
Institute for Theoretical Physics,
University of California,
Santa Barbara, CA 93106-4030; {\tt  wati@itp.ucsb.edu}}\\
{Center for Theoretical Physics} \\
{MIT, Bldg.  6-306} \\
{Cambridge, MA 02139, U.S.A.} \\
{\tt wati@mit.edu}}

\abstract{We describe projection operators in the matter sector of
Witten's cubic string field theory using modes on the right and left
halves of the string.  These projection operators represent a step
towards an analytic solution of the equations of motion of the full
string field theory, and can be used to construct D$p$-brane solutions
of the string field theory when the BRST operator $Q$ is taken to be
pure ghost, as suggested in the recent conjecture by Rastelli, Sen and
Zwiebach.  We show that a family of solutions related to the sliver
state are rank one projection operators on the appropriate space of
half-string functionals, and we construct higher rank projection
operators corresponding to configurations of multiple D-branes.}

\preprint{MIT-CTP-3130, NSF-ITP-01-33, hep-th/0105059}
\begin{document}

\baselineskip16pt
\parskip=4pt

\section{Introduction}

In the last 18 months, renewed attention has been paid to Witten's
cubic bosonic open string field theory \cite{Witten-SFT}, following
Sen's conjectures that this formalism can be used to give an analytic
description of D25-brane decay in bosonic string theory
\cite{Sen-universality}.  While much progress has been made towards
proving Sen's conjectures using the level truncation approximation
\cite{level-truncation}, the large B field simplification
\cite{large-B}, background independent string field theory \cite{bi},
and other methods \cite{other} little progress has been made towards
finding a nontrivial analytic solution of the cubic string field
theory equations of motion
\begin{equation}
Q \Phi =  -\Phi \star \Phi\,.
\label{eq:SFT-EOM}
\end{equation}

It was pointed out by Witten in \cite{Witten-SFT} that the string
field star algebra is roughly describable as the algebra of operators
acting on a space of states associated with a half string.  In such a
formalism, the dependence of the string field on the degrees of
freedom living on the right and left halves of the string are
separated, so the string field can be written in the form $\Psi[l,
r]$.  In the operator formalism the star product, which can be written
as $(\Psi \star \Phi)[l, r] = \int {\cal D}x \Psi[l, x] \Phi[x, r]$
simply becomes multiplication of operators.  Making such an operator
formalism for the string field precise is a difficult problem which
has not yet been solved.  Midpoint insertions in the ghost sector and
the structure of the BRST operator present particular complications.
In this paper we use a formalism of this type to perform calculations
in the matter sector of the string field theory, where there are fewer
subtleties involved.  We describe states in the string field theory in
terms of modes on the left and right halves of a split string.  We use
this formalism to identify solutions of the projection equation
\begin{equation}
\Psi_m = \Psi_m \star \Psi_m
\label{eq:matter-EOM}
\end{equation}
for string fields in the matter sector of the string field theory in
26-dimensional flat space.  This work represents a first step towards
developing a split string formalism for the full string field theory
which could be used to rewrite and perhaps to solve the full string
field theory equation of motion (\ref{eq:SFT-EOM}).  The techniques we
use in this paper can be extended to describe the ghost sector of the
theory, although there are a number of additional complications in
this case; the details of the analysis in the ghost sector will
be described in a later publication \cite{Gross-Taylor-II}.

Recently it was suggested by Rastelli, Sen and Zwiebach (henceforth
RSZ) that in the locally stable vacuum in which the space-filling
D25-brane has been annihilated, it may be possible to describe the
shifted string field theory using a BRST operator which is pure ghost
\cite{rsz}.  With a pure ghost BRST operator of this kind, the
equations of motion (\ref{eq:SFT-EOM}) can be factorized into a ghost
part $\Psi_g$ satisfying (\ref{eq:SFT-EOM}) and a matter part $\Psi_m$
satisfying (\ref{eq:matter-EOM}), when the full string field takes the
form $\Phi = \Psi_m \otimes \Psi_g$.  This factorization was used in
\cite{rsz-2} to identify a certain solution of (\ref{eq:matter-EOM})
found in \cite{Kostelecky-Potting,Rastelli-Zwiebach} as the matter
part of a D25-brane solution.  Even if the RSZ conjecture is not true,
the BRST operators these authors suggest define a new interesting
class of string field theories.  In \cite{rsz}, a particular class of 
candidates
for the pure ghost $Q$ were suggested, namely the operators $Q = \sum
a_n (c_n+ (-1)^nc_{-n})$.  These operators all have trivial cohomology
and satisfy $Q^2 = 0$, both properties which are expected of the BRST
operator in the closed string vacuum.  In addition to these
properties, there are two additional properties which should be
expected of the BRST operator: first, that $\int Q \Psi = 0 \; \forall
\Psi$ (equivalent to the condition $QI = 0$ where $I$ is the identity
of the string field star algebra), and second that $Q$ act as a
derivation of the star algebra.  Not all the operators in the class
suggested by RSZ satisfy both of these conditions, which may give some
clues as to the correct choice of operator.

In the half-string formalism we consider here, the solutions of
(\ref{eq:matter-EOM}) are simply projection operators onto half-string
states, which can be described as functionals of the coordinate fields
on the half string.  A simple class of rank one projection operators
take the form
\begin{equation}
e^{-\frac{1}{2} l \cdot M \cdot l-\frac{1}{2}r \cdot M \cdot r}
\label{eq:projection-functional1}
\end{equation}
where $l, r$ describe the degrees of freedom of the left and right
halves of the string.  We show that the D-instanton state related to
the sliver state found in
\cite{Kostelecky-Potting,Rastelli-Zwiebach,rsz-2} is a projection
operator of this form.  We show that
this state can be modified in various ways to produce other projection
operators describing the D-instanton and higher-dimensional
D$p$-branes.  Rank $r$ projection operators given by the sum of $r$
orthogonal rank one projection operators
can be identified with
solutions of (\ref{eq:matter-EOM}) describing $r$ D-branes.

In Section 2 we develop the half-string formalism and describe the
structure of the cubic string field theory in this language.  In
Section 3 we describe projection operators using half-string degrees
of freedom.  We derive a condition which must be satisfied by a
Gaussian state in the string Fock space for it to have the form of a
rank one projector onto a Gaussian half-string state.  We use the
D-instanton state, which is a rank one projection of this form, to
construct a class of projectors describing D$p$-branes of
arbitrary dimension, and we discuss the construction of higher rank
projectors corresponding to multiple D-branes.  Section 4 contains
some concluding remarks.  The Appendix contains a proof
that the D-instanton state is a rank one projection.  After this work
was completed we learned of related work by Rastelli, Sen and Zwiebach
\cite{rsz-3}.

\section{Half-string formalism}

In this section we develop the formalism to describe string states in
terms of half-string degrees of freedom.  This formalism was
implicitly used to construct the three-string vertex operator in
\cite{Gross-Jevicki-12}.  A related formalism was further developed in
\cite{bcnt,Abdurrahman-Bordes}.  We will only describe the matter
fields $X^i, i \in \{0, 1, \ldots, 25\}$ here; a similar approach can
be used for the ghost fields, which will be presented in detail
elsewhere.  For the most part, we will use the notation and
conventions of \cite{Gross-Jevicki-12}.

\subsection{Matter fields}

The bosonic string has 26 matter fields $X^i$.
We can expand each of these fields in modes through
\begin{equation}
x(\s) =  x_0+ \sqrt{2}\sum_{n=1}^\infty{x_n\cos(n\s)}\,.
\quad
\label{eq:x}
\end{equation}
(We drop spatial indices in most formulae for clarity.)  We can
relate the modes in (\ref{eq:x}) to creation and annihilation
operators through
\begin{eqnarray}
x_n  =  {i \over  {\sqrt{2n}} }
\left(a_n-a^\dagger_n\right) & \hspace{0.5in} &
           p_n =-i
{\p \over \p x_n} =  \sqrt{ n\over 2 } \left(a_n+a^\dagger_n\right)
        \label{eq:xp}\\
a_n = -i \left(  \sqrt{n\over 2} x_n + \frac{1}{ \sqrt{2n}}
\frac{\partial}{ \partial x_n}  \right) & &
a^{\dagger}_n = i \left(  \sqrt{ n\over 2} x_n  - \frac{1}{ \sqrt{2n}}
\frac{\partial}{ \partial x_n}  \right) \, ,\nonumber
\end{eqnarray}
for $n \neq 0$, and through
\begin{eqnarray}
x_0  =  {i \over  {{2}} }
\left(a_0-a^\dagger_0\right) & \hspace{0.5in} &
           p_0 =-i
{\p \over \p x_0} =   \left(a_0+a^\dagger_0\right)
        \label{eq:xp0}\\
a_0 = -i \left(   x_0 + \frac{1}{2}
\frac{\partial}{ \partial x_0}  \right) & &
a^{\dagger}_0 = i \left(   x_0  - \frac{1}{ 2}
\frac{\partial}{ \partial x_0}  \right)\, ,\nonumber
\end{eqnarray}
for the zero modes.
We write
\begin{eqnarray}
\quad |x)  & =&{i\over \sqrt{2}}E[|a)-|a^\dagger)] \\
        \quad |p) & = &
           {1\over \sqrt{2}E} [|a)+|a^\dagger)]\, , \nonumber
\end{eqnarray}
where
\begin{equation}
[\an,a^\dagger_m]=\delta_{nm}, \quad \quad    E^{-1}_{nm}=
\delta_{nm}\sqrt{n}+\delta_{n0}\delta_{m0} \sqrt{2} \, .
\label{eq:can}
\end{equation}
A state in Fock space can be described as a linear combination of Fock
space states of the form $|\Psi\rangle = \prod{{a_1^{\dagger}}^{i_1}
{a_2^{\dagger}}^{i_2}
\cdots {a_n^{\dagger}}^{i_n}\cdots} \vac   $, or as a functional on
the space of
string configurations
\begin{eqnarray}
\Psi\left[x(\s)\right] & = &
\langle x (\sigma) | \Psi \rangle
\label{eq:position}
\\& = & \langle\Omega|\exp\sum_n\left[- \frac{1}{2}
nx_nx_n -x_0^2
+i\sqrt{2n}\an x_n
+ 2ia_0x_0 +\half\an \an \right]|\Psi\rangle  \, . \nonumber
\end{eqnarray}
where
\begin{equation}
|x(\s)\rangle= \exp\left[-{1\over 2}(x|E^{-2}|x)- i(\ad
|\sqrt{2}E^{-1}|x)
  +\half(\ad|\ad)
           \, \right]\vac\,.
\label{eq:xf}
\end{equation}

\subsection{Elements of string field theory}

The string field $\Psi$ is a functional, which can be expressed as
above either in Fock space language or as $\Psi[\{x_n\}]$ in terms of
the string modes as in (\ref{eq:position}).

The star product of two string fields $\Psi_1,
\Psi_2$ can be written  as the functional \cite{Witten-SFT}
\[
        \left(\P_1 \star  \P_2\right) [z(\s)]
        \equiv\int  \P_1 [x(\s)] \P_2 [y(\s)]\prod_{{\pi\over 2} \leq
\s \leq \pi}
\delta[x(\s)-y(\pi-\s)]\prod_{{\pi\over 2} \leq \s \leq \pi}dx(\s)
\prod_{{0} \leq \s \leq {\pi\over 2}} dy(\s) \nonumber
\]
\begin{eqnarray}
z(\s) & = &x(\s) \quad {\rm for} \quad {0} \leq \s \leq {\pi\over 2}\, ,
\label{eq:mult}\\
z(\s) & = &y(\s)\quad {\rm for} \quad   {\pi\over 2} \leq \s \leq \pi\, .
\nonumber
\end{eqnarray}
While the infinite product of delta functions over a finite interval
in (\ref{eq:mult}) may seem to only have formal meaning, this
expression is
given a precise meaning  in terms of Fourier modes on the
string.  This approach was used in \cite{Gross-Jevicki-12} to
construct a Fock space representation of the 3-string vertex.

The identity with respect to the star product $\star$ is the state
\begin{equation}
I[x(\s)]\equiv \prod_{{0} \leq \s \leq {\pi\over 2}}
\delta[x(\s)-x(\pi-\s)]=  \prod_n \delta(x_{2n+1})\, .
\label{eq:identity}
\end{equation}
This state can be written in Fock space language as \cite{Gross-Jevicki-12}
\begin{equation}
|I\rangle=\exp\left[-\half(\ad|C|\ad)\right]\vac =
\exp\left[-\half\sum_{n=0}^\infty
   (-1)^n{a^\dagger_n}{a^\dagger_n} |\right]\vac \,  ,
\label{eq:idenf}
\end{equation}
where C is given by
\begin{equation}
C_{nm}= \d_{nm}(-1)^n \, .
\label{eq:C}
\end{equation}

The integral of a string field $\Psi$ is defined by
\begin{equation}
\int \P =\langle\P | I   \rangle \, ,
\label{eq:in}
\end{equation}
where $\langle \P |$ is the Hermitian conjugate (not the BPZ dual) of
$| \Psi \rangle$.  For a pair of string fields $\Psi_1, \Psi_2$ the
integral of $\Psi_1 \star  \Psi_2$ is given by
\begin{equation}
          \int \P_1\star \P_2 = \int \P_1[x(\s)] \P_2[y(\s)]
\prod_{{0} \leq \s \leq \pi}
\delta[x(\s)-y(\pi-\s)]dx(\s)dy(\s) .
\label{eq:Integ}
\end{equation}

The string field theory action (including ghosts) is given by
\begin{equation}
        S=- \int\left[\frac{1}{2}  \P\star Q\P +
\frac{1}{3}  \P \star  \P \star \P \right]\,.
\label{eq:action}
\end{equation}
The equation of motion from this action is
\begin{equation}
Q \Psi = -\Psi \star \Psi.
\label{eq:SFT-EOM-2}
\end{equation}
The standard string BRST operator $Q$ mixes the matter and ghost
sectors, so that solving (\ref{eq:SFT-EOM-2}) involves a highly
nontrivial interplay between the matter and ghost sectors.  As
mentioned above, in the RSZ model the operator $Q$ is taken to be pure
ghost.  For such a BRST operator,  and for states of the form $\Psi =
\Psi_m \otimes \Psi_g$, the equations of motion  factorize as
discussed in the Introduction and in the matter sector we simply
need to solve the projection equation
\begin{equation}
\Psi_m = \Psi_m \star \Psi_m\,.
\label{eq:projection-equation}
\end{equation}
Even if the conjecture of Rastelli, Sen and Zwiebach is incorrect, it
is still of  interest to find solutions to
(\ref{eq:projection-equation}).  Solving equations of this form has
led to substantial insights into the form of solitons describing
lower-dimensional D$p$-branes in the large B limit and in
noncommutative geometry (see \cite{Harvey-lectures} for a review and
further references).  It seems likely that developing a better
understanding of solutions to (\ref{eq:SFT-EOM-2}) and
(\ref{eq:projection-equation}) separately in the ghost and matter
sectors will lead to better tools for attacking the problem of solving
the full equation of motion in the original theory where $Q$ couples
the matter and ghost sectors.  In the remainder of this paper, we
focus on finding solutions to (\ref{eq:projection-equation}) in the
matter sector.

\subsection{Splitting the string}
\label{sec:splitting}

In this section we describe the transformations needed to rewrite the
string field in terms of modes on the right and left parts of the
string.  We break up the string coordinate $x(\s)$, which satisfies
Neumann boundary conditions at $\s =\ 0 , \pi$, into its left and
right pieces, according to
\begin{equation}
l(\s)=x(\s), \quad r (\s) = x (\pi -\s)
\;\;\;\;\; \quad{\rm for}\quad {{0} \leq \s \leq {\pi\over 2}} \ ,
\label{eq:defxrl}
\end{equation}
where $l(\s)$ and $r(\s)$ obey Neumann boundary
conditions at $\s =0$ and Dirichlet boundary conditions at $\s=\pi/2$.
We can perform a separate mode expansion on the left and right
pieces of the string
\begin{eqnarray}
           l(\s)&=&\sqrt{2}\sum_{n=0}^\infty
   l_{2n+1}
\cos(2n+1)\s \ ,
\label{eq:defrl2}\\
             r( \s)&= &\sqrt{2}\sum_{n=0}^\infty
   r_{2n+1}
\cos(2n+1)\s \ .\label{eq:r-expansion}
\end{eqnarray}
Note that the functions $l (\sigma), r (\sigma)$ defined
through (\ref{eq:defrl2}, \ref{eq:r-expansion}) need not vanish at the
point
$\s=\pi/2$, since for example if $
l_{2n+1}={2\sqrt{2}(-1)^n\over(2n+1)\pi}a +o(1/n)$, then using
$$\sum_n{4(-1)^n\over(2n+1)\pi} \cos(2n+1)\s = 1 \quad{\rm for}\quad
{{0} < \s < {\pi\over 2}},$$ we have $l(\pi/2)= a$.  If
$x (\sigma)$ is a well-behaved (smooth) function then $l
(\sigma)$, $r (\sigma)$ defined through (\ref{eq:defxrl}) will take
the value $x (\pi/2)$
at $\sigma = \pi/2$.

Using
\begin{equation}
\int_0^{\pi\over 2} d\s \cos2n\s\cos(2m+1)\s=
\left\{
\begin{array}{l}
{\pi\over
4}X_{2n,2m+1}\, \;\;\;\;\; ({\rm for} \,\, n\neq 0)\\
{\pi\over
2\sqrt{2}}X_{0,2m+1}\, \;\;\;\;\; ({\rm for} \,\, n= 0)\, ,
\end{array}\right.
\label{eq:cosint}
\end{equation}
where
\begin{eqnarray}
X_{2k+1,2n}=X_{2n,2k+1} & = & { 4(-1)^{k+n}(2k+1)\over
\pi\left({(2k+1)^2-4n^2}\right)} \;\;\;  \quad  (n \neq 0)\, , \label{eq:X}\\
\quad X_{ 2k+1,0} =X_{0,2k+1} & = & 
{  2 \sqrt{2}(-1)^{k}\over \pi{(2k+1)}}\, ,\nonumber
\end{eqnarray}
we can relate the full-string modes to the half-string modes through
\begin{eqnarray}
        x_{2n+1} &= &\half\left(l_{2n+1}-r_{2n+1}\right)  \, ,
\label{eq:xrel}\\
x_{2n}
  & = &  \half
\sum_{k=0}^{\infty}X_{2n,2k+1}\left(l_{2k+1}+r_{2k+1}\right)\, .
\nonumber
\end{eqnarray}

The matrix
\begin{equation}
X \equiv  \pmatrix{  0 & X_{2k+1,2n} \cr
         X_{2n, 2k+1} & 0  \cr }  \,
\label{eq:matX}
\end{equation}
is symmetric and orthogonal: \ \  $X=X^T=X^{-1}$.
Furthermore, we have
\begin{equation}
\sum_{n=1}^{\infty}X_{2k+1,2n}(-1)^n=-{1\over
\sqrt{2}}X_{2k+1,0}, \quad  \quad
\quad
\sum_{n=0}^{\infty}X_{2k+1,2n}(2n)^2X_{ 2n,2k'+1}=(2k+1)^2\d_{kk'}\, .
\label{eq:relX}
\end{equation}
Note that the matrix $X$ defined in (\ref{eq:matX}) differs from the
equivalent $X$ defined in \cite{Gross-Jevicki-12} by a factor of $i$.

We can invert (\ref{eq:xrel}) to derive
\begin{eqnarray}
l_{2k+1}   &=
&x_{2k+1}+\sum_{n=0}^{\infty}X_{2k+1,2n}
x_{2n}  \, ,
\label{eq:xrel2}\\
r_{2k+1} & = &
-x_{2k+1}+\sum_{n=0}^{\infty}X_{2k+1,2n}
x_{2n}  \, .
\nonumber
\end{eqnarray}

Note that care most be taken when using these formulae to deal with
functions which do not satisfy $\sum_{n} n x_n^2 < \infty, \sum_{k} k
(l_{2k + 1}^2 + r_{2k + 1}^2) < \infty$.  If, for example, we take
$l_{2k + 1} + r_{2k + 1} = X_{2k + 1, 0} = 4(-1)^k/\pi (2k + 1)$, then
we must take care with the order of summation when evaluating $x
(\pi/2) = \sum (-1) x_{2n}$ since
\begin{equation}
{\sqrt{2}-1\over  \sqrt{2}} = \sum_{k = 0}^{\infty} \left[
   (\sum_{n = 0}^{\infty} (-1)^n  X_{2n, 2k + 1})
X_{2k + 1, 0} \right] \neq
\sum_{n = 0}^{\infty} \left[
   (-1)^n \sum_{k = 0}^{\infty} \left(  X_{2n, 2k + 1}
X_{2k + 1, 0} \right) \right] = 1\, .
\label{eq:subtle-summation}
\end{equation}

\subsection{Operator representation of  star algebra}
\label{sec:operator}

The full-string position basis states $| x (\sigma) \rangle$ defined
through (\ref{eq:xf}) can be related to a half-string position basis
through
\begin{equation}
| x (\sigma) \rangle = |\{x_n\} \rangle
= | l (\sigma), r (\sigma) \rangle
= |\{l_{2k + 1}\},\{r_{2k + 1}\} \rangle\, ,
\end{equation}
using (\ref{eq:defxrl}, \ref{eq:xrel}).

For every state  $|\P \rangle $ in the Hilbert space  there is an
associated functional $\Psi[x (\sigma)]$  given by
\begin{equation}
   \P[x(\s)] = \P[l(\s),r(\s)] = \langle l (\sigma), r (\sigma) | \Psi
   \rangle\,.
\label{eq:functional}
\end{equation}
We  can associate the functional (\ref{eq:functional}) with an
operator $\hat{\Psi}$ on the space of functionals of the half-string
degrees of freedom, which we write formally as
\begin{equation}
\Psi[x (\sigma)]
   \Rightarrow   \hat \P = \int\CD l\CD r
\;|l\rangle \P [l,r]  \langle r| \ ,
\label{eq:map}
\end{equation}
or equivalently:
\begin{equation}
\langle x(\s)|\P \rangle = \langle l | \hat \P
|r\rangle  \, .
\label{eq:mapp}
\end{equation}
These functionals are given precise meaning by using the left and
right half-string modes described in Subsection \ref{sec:splitting}.

In the language of operators,
integration of the string field corresponds to taking the trace of the
operator
\begin{eqnarray}
\int \P &\Rightarrow  & \Tr [\hat\Psi] \,.
\label{eq:rel1}
\end{eqnarray}
The star product $\star$ is equivalent to the multiplication of
operators acting on half-string functionals
\begin{equation}
            \P_1\star \P_2 \Rightarrow   \hat \P_1 \hat \P_2 \, .
\label{eq:rel2}
\end{equation}
The identity of the star algebra becomes the identity operator on
half-string states
\begin{equation}
\quad I[x(\s)]= \prod_{{0} \leq \s \leq {\pi\over 2}}
\delta[l(\s)-r( \s)]  \Rightarrow   \hat I =
\int  \CD l(\s)|l(\s)\rangle \langle l(\s)| \ .
\label{eq:id}
\end{equation}

For the purposes of this paper we primarily think of the functional
$\Psi[l, r]$ associated with the Fock space state $| \Psi \rangle$ as
a single operator acting on the space of functionals on the
half string.  One way of describing the action of this operator is to
explicitly separate out the midpoint $x = x (\pi/2) = \sum_{n}
(-1)^nx_{2n}$.  We can modify (\ref{eq:xrel}) and (\ref{eq:xrel2}) by
replacing $x_0 \rightarrow x_0-x$, so that $l (\sigma), r (\sigma)$
defined through (\ref{eq:defrl2}, \ref{eq:r-expansion}) satisfy $l
(\pi/2) = r (\pi/2) = 0$ when $x (\sigma)$ is smooth.  The functional
$\Psi[l, r]$ can then be described as an operator-valued function of
$x$, $\hat{\Psi} (x)$, where the operator at each point in space-time
acts on a separate copy of the usual Hilbert space associated with a
string having Neumann boundary conditions on one end and Dirichlet
boundary conditions on the other.  In transforming to half-string
variables, only $x_0$ depends upon the midpoint value $x$, so that a
Fock space state with well-defined momentum $p_0$ becomes an
operator-valued functional of the form $e^{ip_0 x} \hat{\Psi}$ where
$\hat{\Psi}$ acts on the ND string Fock space.  This construction can
be generalized to an arbitrary space-time manifold ${\cal M}$, where
$\hat{\Psi} (x)$ becomes an operator-valued function on ${\cal M}$.  A
particularly simple class of operators are those which are independent
of the zero mode $x_0$, and hence of the midpoint $x$.  These
operators can be thought of simply as operators acting on the usual
Hilbert space of a string having Neumann boundary conditions on one
end and Dirichlet boundary conditions on the other end.  While the
point of view where the midpoint is treated separately becomes helpful
in addressing certain problems in the ghost sector of the theory, we
will not treat the midpoint separately in most of the equations
in this paper.  We will mention this point of view again briefly when
discussing some points where this perspective is helpful.

\subsection{Squeezed states}

In this section we collect some formulae which will be useful in
discussing projection operators in the split string formalism.

A particularly nice class of states are squeezed states of the form
\begin{equation}
\exp \left({-\frac{1}{2}a^{\dagger}_nS_{nm}a^{\dagger}_m}\right)
| \Omega \rangle\,.
\end{equation}
Up to an overall constant, such a state is described by the functional
\begin{equation}
\exp \left\{{-\frac{1}{2}\left(x | E^{-1} \left[ \frac{1 - S}{1 + S} \right]
E^{-1} | x\right)}\right\}
  \,.
\label{eq:squeezed-functional}
\end{equation}

This relationship can easily be inverted.  The functional
\begin{equation}
\exp \left({-\frac{1}{2}\left(x | E^{-1}   D
E^{-1} | x\right)}\right)
  \,
\end{equation}
corresponds (up to an overall constant) to the state
\begin{equation}
\exp \left({-\frac{1}{2}\left(a^{\dagger} |
\frac{1- D}{1 +  D} |  a^{\dagger} \right)}\right)
| \Omega \rangle\,.
  \label{eq:c-squeezed}
\end{equation}

\section{Projection operators}

We will now use the half-string formalism developed in the previous
section to describe solutions to the projection equation
\begin{equation}
\Psi = \Psi \star \Psi\,.
\label{eq:matter-equation}
\end{equation}
Such solutions may be useful for finding an analytic solution to the
full SFT equations, as discussed for example by Kostelecky and Potting
\cite{Kostelecky-Potting}.  If the pure ghost Ansatz for the BRST
operator in the vacuum suggested by Rastelli, Sen and Zwiebach
\cite{rsz} is correct, solutions to (\ref{eq:matter-equation}) will
correspond to D-branes condensing out of the vacuum.

Using the operator description (\ref{eq:map}) of string states,
solutions to (\ref{eq:matter-equation}) simply correspond to
projection operators on the space of functionals on the half string.
Such projection operators can be classified by their rank $r$.  The
identity $\hat{I}$ is an example of an infinite rank projection
operator.

In Subsection \ref{sec:rank1} we discuss rank one projection operators
on the space of half-string functionals. We formulate conditions on a
Gaussian state in the full string Fock space which must be satisfied
for it to correspond to a rank one projector onto a Gaussian state of
the half string.  In the Appendix we show that the D-instanton state
related to the sliver satisfies these conditions.  Subsection
\ref{sec:others} describes a number of ways in which the D-instanton
state can be used to construct a large number of other
projection operators corresponding to other D-instanton and
higher-dimensional D-brane states, and addresses the question of when
these states are gauge equivalent.  In \ref{sec:multiple} we describe
multiple D-brane configurations corresponding to higher rank
projections.

\subsection{Rank one projection operators}
\label{sec:rank1}

The general form of a  rank one projector is
\begin{equation}
\hat{\Psi} = | \chi \rangle \langle \chi | \, ,
  \label{eq:general-rank1}
\end{equation}
where $\Psi[x (\sigma)]$ takes the factorized form
\begin{equation}
\Psi[l (\sigma), r (\sigma)] = \chi [l (\sigma)] \chi[r (\sigma)]\, ,
  \label{eq:factorized-form}
\end{equation}
and
\begin{equation}
\chi[l (\sigma)] = \langle l (\sigma) | \chi \rangle\,.
\end{equation}

One simple class of rank one projection operators are described by
Gaussian functionals
\begin{equation}
\chi[l (\sigma)] \sim
\exp\left({-\frac{1}{ 2} l_{2k + 1}M_{kj}l_{2j + 1}} \right)\,.
\label{eq:Gaussian-projection}
\end{equation}
Such a projection operator is described in the full string mode basis
by a functional
\begin{equation}
\exp \left({- \frac{1}{2} x_n L_{nm} x_m} \right)\, ,
  \label{eq:projection-functional}
\end{equation}
where
\begin{eqnarray}
L_{2k + 1, 2j + 1} & = &  2M_{kj} \nonumber\\
L_{2n, 2m} & = &  2 X_{2n, 2k + 1} M_{kj} X_{2j + 1, 2n}
\label{eq:projection-form}\\
L_{2 k + 1, 2m} = L_{2n, 2j + 1} &= & 0\,. \nonumber
\end{eqnarray}

Turning this discussion around, we can state the conditions on a state
of the form (\ref{eq:projection-functional}) which are necessary for
it to be a projection operator of the form
(\ref{eq:Gaussian-projection}).  To be a projection operator onto a
Gaussian half-string state, it must
be the case that:\\
\vspace{0.04in}

\noindent {\bf (a)} The components $L_{nm}$ must vanish when $n+ m$ is odd.\\
\noindent {\bf  (b)} The nonzero components must satisfy
\begin{equation}
L_{2n, 2m} = X_{2n, 2k + 1} L_{2k + 1, 2j + 1} X_{2j + 1, 2m}\,.
\label{eq:projection-conditions}
\end{equation}
\vspace*{0.06in}

There are some simple functionals which satisfy conditions (a) and (b).
For example, from (\ref{eq:relX}) we see that
\begin{equation}
L_{nm} = \delta_{nm}  \;\;\;\;\; {\rm and} \;\;\;\;\;
L_{nm} = \delta_{nm} n^2
\label{eq:almost}
\end{equation}
satisfy (\ref{eq:projection-conditions}) as well as condition (a).
The corresponding half-string states are given by
(\ref{eq:Gaussian-projection}) where
\begin{equation}
M_{kj} = \frac{1}{2}\delta_{kj}\;\;\;\;\; {\rm and} \;\;\;\;\;
M_{kj} = \frac{1}{2}\delta_{kj} (2k + 1)^2\,.
\end{equation}
There is a further condition, however, which we would like the
state $\Psi$ associated with the projection operator to satisfy, which
is that it should be a normalizable state in the Fock space built on
the ground state whose wavefunction  is
\begin{equation}
\Psi_0[\{x_n\}] \sim\exp \left(-\sum_{n= 1}^{\infty}
  \frac{n}{2}x_n^2 -x_0^2\right)\,.
\end{equation}
This leads us to impose the additional condition that the matrix
$L_{nm}$ behave like $n \delta_{nm}$ for large $n, m$.  If we describe
the projection operator as a squeezed state in the Fock space basis
through
\begin{equation}
| \Psi \rangle =
e^{-\frac{1}{2}a^{\dagger}_nS_{nm}a^{\dagger}_m}| \Omega \rangle\, ,
\label{eq:squeezed}
\end{equation}
then by (\ref{eq:squeezed-functional}) we have
\begin{equation}
L =  E^{-1} \left[\frac{1 - S}{1 + S} \right] E^{-1}\,.
\end{equation}
The Fock space normalizability condition on $L$ then implies the
condition that $S$ have a spectrum with absolute value bounded above
by 1.  If the eigenvalues of $S$ are $s_i$ we must have $
\sum_{i}s_i^2 < \infty$.  This normalizability condition is not
satisfied by either of the states defined in (\ref{eq:almost}).

In this discussion we have not worried about the overall normalization
of states such as (\ref{eq:general-rank1}) which is needed for them to
satisfy the projection equation (\ref{eq:matter-equation}).  Even if a
state $\Psi$ is normalizable in the Fock space and is described by a
functional which can be factorized through (\ref{eq:factorized-form}),
it may be that an infinite or vanishing normalization factor is needed
for (\ref{eq:matter-equation}) to be satisfied.  As discussed in
\cite{rsz-2}, however, there is an undetermined constant which appears
when we separate the matter and ghost sectors.  Actually, we only
really require that the matter field satisfy
\begin{equation}
\Psi_m = \kappa \Psi_m \star \Psi_m\, ,
\end{equation}
where the ghost field satisfies $Q \Psi_g= \kappa^{-1} \Psi_g \star
\Psi_g$ for some $\kappa$.  In this paper we do not worry about the
exact normalization constant $\kappa$, which may even formally be 0 or
$\infty$.

To summarize, we have three conditions which a Gaussian state of the
form (\ref{eq:Gaussian-projection}) must satisfy to be a rank one
projection operator.  The matrix $L_{nm}$ must satisfy conditions (a)
and (b), and the associated squeezed state must have a matrix $S$
whose spectrum is such that (\ref{eq:squeezed}) is a finite norm state
in the Fock space.  It may seem quite difficult to find such a state,
since the condition on the spectrum of $S$ and the relation
(\ref{eq:projection-conditions}) are tricky to balance.  Nonetheless,
it turns out that a class of states related to the sliver state $| \Xi
\rangle$ found in \cite{Kostelecky-Potting,Rastelli-Zwiebach} satisfy
all these conditions.

The sliver state was first constructed as a solution of
(\ref{eq:matter-equation}) by Kostelecky and Potting in
\cite{Kostelecky-Potting}.  These authors constructed this state as a
zero momentum squeezed state of the form (\ref{eq:squeezed}), and gave
a formula for $S_{mn}$ in terms of the coefficients appearing in the
Fock space representation of the cubic interaction vertex describing
the string field star product.  They found numerical evidence that the
eigenvalues of $S$ are smaller than one in absolute value and converge
to zero as needed for the state to be normalizable in the Fock space
\footnote{Note added: It may be, however, that $S$ nonetheless has an
eigenvalue of -1, taking the sliver state slightly outside the Hilbert
space \cite{rsz-4,Sen-personal}}.  In \cite{rsz-2}, Rastelli Sen and
Zwiebach conjectured that the sliver state corresponds to the
space-filling D25-brane in the matter sector of the model they
proposed with pure ghost BRST operator.  They also constructed a
general class of solitonic solutions related to the sliver state but
localized in some subset of space-time dimensions.  They suggested
that these lump solutions correspond to lower-dimensional D$p$-branes,
and demonstrated that these solutions had the correct tensions
relative to the D25-brane solution.

In the Appendix, we show that the D-instanton relative $| \Xi_{-1}
\rangle$ of the sliver state, which is localized in all space-time
dimensions, is a rank one projection operator on the space of
half-string states, and we identify the associated functional in terms
of half-string degrees of freedom.  We expect that all the
higher-dimensional lump solutions, including the space-filling sliver
state can be similarly described as projection operators which can be
written explicitly as functionals of the half-string degrees of
freedom.  In the next subsection we show how the D-instanton state can
be used to construct a range of other rank one projection operators
corresponding to D-instantons and higher-dimensional D-branes.

\subsection{Other projectors from the D-instanton sliver}
\label{sec:others}

{}From the descriptions of half-string projection operators in terms of
full- and half-string modes, we can see that given one projection such
as the D-instanton, it is straightforward to construct a wide class of
other projections.  One way of doing this is to modify a projection
operator defined by an $L$ satisfying (\ref{eq:projection-conditions})
by adding a small shift to $L$ which also satisfies
(\ref{eq:projection-conditions}) but does not violate the
normalizability condition.  For example, if we start with the
D-instanton rank one projector having the $L$ defined through
(\ref{eq:l-instanton}, \ref{eq:s}) in the Appendix, we can define a
projection operator with a different width in space-time through
\begin{equation}
\tilde{L}_{nm} = L_{nm} +  \lambda_{nm}\, ,
\label{eq:l-shift}
\end{equation}
where $\lambda_{00} = \delta$,
\begin{equation}
\lambda_{2k + 1, 2j + 1} = X_{2k + 1, 0} X_{0, 2j + 1} \delta
= \frac{(-1)^{j + k} 8}{ \pi^2 (2k + 1) (2j + 1)} \delta \,,
\end{equation}
and all other elements of $\lambda$ vanish.  The new operator
defined by $\tilde{L}$ is clearly a rank one projection, since
(\ref{eq:projection-conditions}) is still satisfied, and the small
change in $L$ for large $k, j$ should not affect the normalizability
of the state.  By performing this shift with different values of
$\delta$ in each space-time direction, we can construct a large class
of different rank one projection operators corresponding to
D-instantons localized at the origin.  We expect that all these
operators are related to one another by gauge transformations.  A
similar family of (presumably gauge equivalent) projection operators
corresponding to branes with different widths in the transverse
directions was described in \cite{rsz-2} by using a one-parameter
family of creation and annihilation operators to describe the
zero-modes $x_0$.

Starting with the D-instanton, we can modify the projection operator
more dramatically by using a shift of the form (\ref{eq:l-shift}) to
set all coefficients $L_{0n}$ to zero in $p + 1$ of the space-time
dimensions $X^{\parallel}$, shifting the coefficients $L_{2k + 1, 2j +
1}$ as needed to maintain the relation
(\ref{eq:projection-conditions}).  The resulting state is not
normalizable, but this is because the string functional is independent
of $p + 1$ of the zero modes $x_0^\parallel$.  Dividing the norm of
the state by the $p + 1$-dimensional volume of space-time in these
directions gives us a finite normalization for these states.  These
projection operators are really projectors of infinite rank on the
full space of half-string states, but they are rank one on the space
of states independent of $x^\parallel$.  We refer to these as
``transverse rank one'' projection operators.  The projection
operators constructed in this fashion correspond to higher-dimensional
D$p$-brane states in the RSZ model.  It is not clear whether the
functionals describing these states will be identical with the
higher-dimensional solutions found in \cite{Kostelecky-Potting,rsz-2},
but the only difference for geometrically identical brane solutions
should be a nontrivial gauge transformation which may be needed to
move from one description to the other.

Another way to construct new rank one projection operators
given a single projector onto a Gaussian state is to multiply the
state by fields which take the left and right parts of the string
separately into a new state.  For example, given a state described by
a functional
\begin{equation}
e^{-\frac{1}{2}l \cdot M \cdot l-\frac{1}{2}r \cdot M \cdot r}\, ,
\label{eq:basic-functional}
\end{equation}
that satisfies the conditions needed for normalizability in the Fock
space, we can construct a projection of the form
\begin{equation}
l_{2k + 1} e^{-\frac{1}{2}l \cdot M \cdot l-\frac{1}{2}
r \cdot M \cdot r} r_{2k + 1}
\end{equation}
that projects onto an orthogonal state from
(\ref{eq:Gaussian-projection}) for any value of $k$.  In the language of
full-string functionals, such a state would be described by
\begin{equation}
\left[\left(\sum_{n}  x_{2n} X_{2n, 2k + 1} \right)^2 -x_{2k + 1}^2 \right]
\exp \left({- \frac{1}{2} x_n L_{nm} x_m} \right)
\end{equation}
where $L_{nm}$ is given by (\ref{eq:projection-form}).  This state can
be described in the Fock space using (\ref{eq:c-squeezed}) and
(\ref{eq:xp}, \ref{eq:xp0}).  This construction can clearly be
generalized by multiplying (\ref{eq:basic-functional}) by any set of
functions of the form $f_a (\{l_{2k + 1}\})f_a (\{r_{2k + 1}\})$, $a =
1, \ldots, r$ and then orthonormalizing to get $r$ independent rank
one projectors.

We can of course translate or rotate any of the solutions just
described by acting on the zero modes $x_0$ in the appropriate
fashion.  This gives us a variety of ways to modify the single
projection operator we have described explicitly in the Appendix,
giving descriptions of single D$p$-branes of arbitrary dimension and
position in terms of functionals of half-string variables.  In the
next subsection we describe how these transverse rank one projections can be
combined to construct higher rank projection operators corresponding
to multiple D-brane states.  We conclude this subsection with a
discussion of the gauge equivalence of different rank one projectors.

All rank $r$ projection operators on a Hilbert space ${\cal H}$ are
equivalent to one another under conjugation by a unitarity transformation $U$
\begin{equation}
\hat{\Psi}_r = U \hat{\Psi}'_r U^{\dagger}\,.
\end{equation}
In string field theory, such transformations can be realized by
exponentiating gauge transformations with gauge parameter $\Lambda_m
\otimes I_g$ which leave the ghost part of a tensor product state
$\Psi_m \otimes \Psi_g$ unchanged.  Naively, this would suggest that
all the rank one projection operators we have described here should be
gauge equivalent in string field theory.  This is contrary to our
understanding of the physics, which suggests that D$p$-branes located
at different points in space-time should not be gauge equivalent.  (Of
course, a D$p$-brane and a D$q$-brane for $p \neq q$ are not
equivalent since they are described by rank one projection operators
on spaces of functionals independent of different sets of zero modes
$x^\parallel$.)  To understand the resolution of this puzzle regarding
two D$p$-branes with transverse separation, let us
consider a pair of D-instanton states centered at different points $0$
and $y$ in space-time.  These two states can be written in the full
string Fock space as
\begin{equation}
| \Xi_{-1} \rangle, \;\;\;\;\; e^{ y \partial/\partial_{x_0}}
| \Xi_{-1} \rangle\,.
\label{eq:two-instantons}
\end{equation}
In the Appendix we show that $| \Xi_{-1} \rangle$ can be described in
the split string language as a functional $\chi [l] \chi [r]$ where
\begin{equation}
\chi [\{l_{2k + 1}\}] =e^{-l \cdot L \cdot l/4}
\end{equation}
and $L_{2k + 1, 2j + 1}$ is given in (\ref{eq:l-instanton}) in terms
of (\ref{eq:s}).  The
generator of a unitarity transformation which takes a projector $|
\chi \rangle \langle \chi |$ into an orthogonal projector $| \alpha
\rangle \langle \alpha |$ is given by $| \chi \rangle \langle \alpha |
-| \alpha \rangle \langle \chi |$.  Thus, the matter part of the gauge
transformation taking us from the D-instanton centered at the origin
to the D-instanton centered at the space-time point $y$, if it
exists, should be given by
\begin{equation}
\Lambda[l, r] =\chi [l] \chi [r + y] -\chi [l + y] \chi [r].
\end{equation}
Let us consider the first term in this expression.
Translating this back into the language of the full string Fock space,
$\chi[l] \chi[r + y]$ becomes
\begin{equation}
\exp \left(
\frac{y}{2}  \frac{\partial}{ \partial_{x_0}}
- \sum_{k} \frac{y}{2} X_{0, 2k + 1}
   \frac{\partial}{ \partial_{x_{2k + 1}}}
\right)| \Xi_{-1} \rangle\,.
\label{eq:bad-generator}
\end{equation}
While both states in (\ref{eq:two-instantons}) are normalizable states
in the Fock space, the same is not true of (\ref{eq:bad-generator}).
This state has a unit norm but has a vanishing inner product with all
the states in the Fock space.  This result essentially follows from
the fact that the shifts in (\ref{eq:bad-generator}) have the effect
of breaking the string so that the left and right halves of the string
do not have the same value at $\sigma = \pi/2$.  This means that the
resulting functional has vanishing overlap with states in the Fock
space, which are localized around smooth string configurations.  To
see this more concretely, consider the related state
\begin{equation}
\exp \left(
\frac{y}{2}  \frac{\partial}{ \partial_{x_0}}
- \sum_{k} \frac{y}{2} X_{0, 2k + 1}
   \frac{\partial}{ \partial_{x_{2k + 1}}}
\right)| \Omega \rangle\,.
\label{eq:bad-vacuum}
\end{equation}
Rewriting the derivatives in terms of creation and annihilation
operators we have
\begin{equation}
\exp \left[i \frac{y}{2}  (a_0 + a_0^{\dagger})
-i \frac{y}{2}  \sum_{k = 0}^{ \infty}
\frac{(-1)^k 2}{ \pi \sqrt{2k + 1}}  (a_{2k + 1} + a_{2k + 1}^{\dagger})
\right]| \Omega \rangle\,.
\end{equation}
Normal ordering gives us
\begin{equation}
\exp \left[i \frac{y}{2}  a_0^{\dagger}
-i \frac{y}{2}  \sum_{k = 0}^{ \infty}
\frac{(-1)^k 2}{ \pi \sqrt{2k + 1}}   a_{2k + 1}^{\dagger}
-\frac{y^2}{8}  \left( 1 + \sum_{k}\frac{4}{ \pi^2 (2k + 1)}
\right)
\right]| \Omega \rangle\,.
\end{equation}
The infinite negative constant indicates that the overlap of this
state with any state in the Fock space vanishes.  On the other hand,
taking the inner product of this state with itself we see that the
normal ordering constant cancels, so that this state has norm one
(which is expected since it is simply a translation in an infinite
number of modes from the original state).  This shows that the
state does not lie in the usual Fock space.
Since the state $|\Xi_{-1} \rangle$ is in the Fock space, the
same argument shows that (\ref{eq:bad-generator}) is not in the Fock
space.  Since this state is not in the Fock space, it cannot be used
to generate a gauge transformation in the matter sector.  This
explains why the states (\ref{eq:two-instantons}) are not gauge
equivalent.

In this subsection we have constructed a family of rank one projection
operators onto half-string functionals independent of $p + 1$
space-time directions $x^\parallel$.  These projection operators
correspond in the BPZ model to matter sector descriptions of
D$p$-branes of arbitrary dimension, position and orientation.  In the
following subsection we discuss how these projections may be combined
to form higher rank projections describing arbitrary multiple D-brane
configurations.

\subsection{Higher rank projections and multiple D-branes}
\label{sec:multiple}

In general,  projection operators on the space of
half-string functionals
\begin{equation}
\hat{P}_1 = | \chi \rangle \langle \chi |\, ,
\end{equation}
that are rank one on the space of functionals independent of some set
of space-time directions $x^\parallel$
will correspond to single D-brane configurations, when the matter
sector of the string field theory is decoupled as in the RSZ model
\cite{rsz}.  We have identified a particular class of rank one
projection operators as functionals on the full string satisfying
certain conditions.

We can combine $r$ mutually orthogonal transverse rank one projection
operators into a projection operator of
the form
\begin{equation}
\hat{P}_r = | \chi_1 \rangle \langle \chi_1 |+
  | \chi_2 \rangle \langle \chi_2 |+ \cdots + |
\chi_r \rangle \langle \chi_r |\, ,
\label{eq:r-projection}
\end{equation}
where $\langle \chi_i | \chi_j \rangle\sim\delta_{ij}$.  These
operators correspond to configurations of $r$ distinct D-branes, which
may be coincident or separated in transverse space-time dimensions.
In the previous subsection we described a number of ways of constructing
projection operators describing single D$p$-branes of different
dimensions.  Other constructions of single D$p$-brane projection
operators were given in \cite{rsz-2}.  By choosing a set of $r$
orthogonal projectors constructed in any of these ways, we can
construct a projector through (\ref{eq:r-projection}) describing a
system of $r$ D-branes.  This allows us in principle to construct any
multiple D-brane configuration as a projection operator in the matter
sector of the string field theory.

Just as some pairs of rank one projectors are gauge equivalent and
others, which correspond to physically distinct D-branes, are not
gauge equivalent, some rank $r$ projectors are gauge equivalent and
others are not.  An example of interest is given by a set of $r$
orthogonal transverse rank one projectors corresponding to zero
momentum states in the full string Fock space.  As discussed in
subsection (\ref{sec:operator}), when we consider the state $\Psi$ as
a function of space-time taking values in the space of operators on
the ND string Hilbert space ${\cal H}$, zero momentum states are
identified with single operators on ${\cal H}$, so the $r$ projectors
are each rank one when considered as operators on ${\cal H}$.  We
expect all states describing $r$ D25-branes to be gauge equivalent,
which corresponds to the fact that any rank $r$ projector on ${\cal
H}$ can be related to any other rank $r$ projector by a unitarity
transformation on ${\cal H}$ which can be realized as a gauge
transformation in string field theory parameterized by a zero momentum
field.

Let us now describe the form of the potential around a ``transverse
rank $r$'' projection operator describing $r$ coincident D-branes, to
verify that there are $r^2$ tachyonic modes as we expect from
perturbative string theory.  If we write a basis for the space of
half-string functionals independent of some space-time coordinates
$x^\parallel$ as $| \chi_i \rangle, i \in {\bb Z}$, then for a state
described by an operator which is a sum of projections of the form
\begin{equation}
\hat{\Psi} = \sum_{i}| \chi_i \rangle \psi_i \langle \chi_i |\, ,
\label{eq:projection-sum}
\end{equation}
the string field potential will be (removing an overall factor of
$\int \Psi_g \star Q\Psi_g+ \Psi_g \star \Psi_g \star \Psi_g$)
\begin{equation}
V = {\rm Tr}\; \left[\frac{1}{2} \hat{\Psi}^2 -\frac{1}{3}
\hat{\Psi}^3 \right]
= \sum_{i} \frac{1}{2}  \psi_i^2 -\frac{1}{3} \psi_i^3 \, .
\end{equation}
The state (\ref{eq:projection-sum}) solves the equations of motion if
$\psi_i \in\{0, 1\} \; \forall i$.  The rank $r$ projections have $r$
nonzero values of $\psi_i$ and can be written in the form
(\ref{eq:r-projection}) by a relabeling of the basis.  If we expand
the full $x^\parallel$-independent string field in the split string
language
\begin{equation}
\hat{\Psi} = \sum_{i, j}
|\chi_i \rangle \psi_{ij} \langle \chi_j |\, ,
\end{equation}
then we can calculate the quadratic terms in the action around the
rank $r$ projection (\ref{eq:r-projection}).  These terms are given by
\begin{equation}
\sum_{i = 1}^\infty \sum_{j = 1}^{\infty}  \frac{1}{2} \psi_{ij} \psi_{ji}  -
\sum_{i = 1}^{r}  \sum_{j = 1}^{\infty}    \psi_{ij} \psi_{ji}\,.
\end{equation}
For fields $\Psi_{ij}$ with $i, j > r$ we have a positive quadratic
term.  These are massive degrees of freedom.  For fields with one
index $ \leq r$ and the other $> r$ we have vanishing quadratic terms.
These fields correspond to gauge degrees of freedom which rotate one
form of the projection operator into another, as discussed in the
previous subsection.  The $N^2$ fields with $1 \leq i, j \leq r$ all
have negative quadratic terms.  These correspond to the $r^2$
tachyonic open string fields on a system of $r$ unstable D25-branes.
Note that if the branes we considered here were not coincident, the
operators $| \chi_i \rangle \langle \chi_j |, i \neq j$ would not lie
in the full string Hilbert space, like the gauge transformation
parameter (\ref{eq:bad-generator}), so we would only have $r$
tachyonic modes, as we expect for $r$ separated D$p$-branes.

\section{Conclusions}

In this paper we have shown that in the matter sector of Witten's
cubic open string field theory it is possible to describe string
fields as operators on the space of functionals associated with the
degrees of freedom on a half string.  These functionals were described
in terms of half-string oscillator modes.
We identified rank $r$
projection operators as configurations of $r$ D-instantons in the
Rastelli-Sen-Zwiebach model where the BRST operator $Q$ is taken to be
pure ghost, and rank $r$ projection operators on functionals
independent of $p + 1$ zero modes as configurations of $r$
D$p$-branes.  We showed that rank one projection operators onto
Gaussian states of the half-string could be described as Gaussian
states of the full string satisfying certain conditions.  We showed
that a particular D-instanton state constructed by Rastelli, Sen, and
Zwiebach satisfies these conditions and can be identified as a rank
one projection.  We showed how this state can be modified to describe
a D$p$-brane of arbitrary dimension, and that multiple such states can
be added to produce an arbitrary configuration of D-branes. Finally,
we showed that the number of tachyonic degrees of freedom around a
rank $r$ projector describing $r$ coincident D-branes is $r^2$, in
agreement with the expectation from perturbative string theory.

In order to really understand the structure of the vacuum in cubic
string field theory, either with the standard perturbative string BRST
operator $Q$ or with a ghost BRST operator as suggested by RSZ, it is
clearly essential to develop an analogous analytic approach to
describing the ghost sector of the theory.  The most difficult part of
representing the ghost sector in terms of half-string degrees of
freedom is the representation of the ghost number one BRST operator in
terms of the half-string degrees of freedom.  In a sequel to this
paper we will treat this problem and describe the ghost sector of the
theory in a similar fashion to the way in which we have described the
matter sector of the theory in this work.

\appendix

\section{Appendix}

In this appendix we prove that a particular solution of the matter
string field equation of motion $\Psi = \Psi \star \Psi$ which was
found in \cite{Kostelecky-Potting,rsz-2} is a rank one projection
operator onto a Gaussian state on the half string.  In
\cite{Kostelecky-Potting}, Kostelecky and Potting found a
zero-momentum solution of the equations of motion by making the
assumption that the solution took the form of a particular kind of
squeezed state and then by solving the equation of motion explicitly
using the mode representation of the three string vertex.  Later,
Rastelli, Sen and Zwiebach identified the state found by Kostelecky
and Potting as the sliver state found in \cite{Rastelli-Zwiebach}.
These authors also found a more general class of solutions related to
the sliver state but with Gaussian dependence on space-time zero
modes.  They identified these states as D$p$-branes with $p < 25$, and
showed that the ratio of tensions between a pair of D$p$-branes of
different dimensions is correctly reproduced by these Gaussian states.
The precise state we consider here is a squeezed state which is
localized in all space-time dimensions, namely the D-instanton lump
solution constructed by RSZ.  We show that this state is a rank one
projection on the half string using the formalism developed in the
text.  The connection of other higher-dimensional D$p$-brane solutions
to this D-instanton is discussed in the main text.

We begin by recalling some useful formulae from
\cite{Gross-Jevicki-12} and the definition of the D-instanton state
following \cite{Kostelecky-Potting,rsz-2}.  The D-instanton state is a
state of the form
\begin{equation}
| \Xi_{-1} \rangle=
e^{-\frac{1}{2}a^{\dagger}_nS_{nm}a^{\dagger}_m}| \Omega \rangle\,\, ,
\label{eq:squeezed-2}
\end{equation}
with
\begin{equation}
S = CT\, ,
\label{eq:s}
\end{equation}
where $C$ is defined in (\ref{eq:C}) and
\begin{equation}
T = \frac{1}{2Z}  \left( 1 + Z - \sqrt{(1 + 3Z) (1-Z)} \right)\,.
\label{eq:t}
\end{equation}
(Note that in \cite{Kostelecky-Potting,rsz-2}, $Z$ is called $X$, but
this conflicts with the notation used here and in
\cite{Gross-Jevicki-12} for $X$).  The matrix $Z$ is given by
\begin{equation}
Z = CV\, ,
\label{eq:z}
\end{equation}
where $V$ is the matrix $U^{11}$ appearing in \cite{Gross-Jevicki-12}
in the cubic string vertex constructed using creation-annihilation
operators $a_0, a_0^{\dagger}$.  (Note that there are several
different notations used in the literature for the matrices $U, V$ we
consider here.  The matrices we call $U, V$ are the same as $U,
U^{11}$ in \cite{Gross-Jevicki-12}, $U', V'$ in
\cite{Kostelecky-Potting}, and $U^{gj}, V'$ in \cite{rsz-2}.)  The
matrix $V$ can in turn be related to the change of basis matrix $X$
defined in (\ref{eq:matX}) by writing
\begin{equation}
V =\frac{1}{3} \left( C + U + \bar{U} \right)
\label{eq:v}
\end{equation}
where
\begin{equation}
U = \left(2-EYE^{-1} + E^{-1} Y E\right) \left[
E Y E^{-1} + E^{-1} Y E \right]^{-1},
\label{eq:u}
\end{equation}
$E$ is defined in (\ref{eq:can}),
and
\begin{equation}
Y = -\frac{1}{2}C - \frac{\sqrt{3}}{2}  i C X\,.
\label{eq:y}
\end{equation}

{}From these definitions we can derive some useful relations.  We have
\begin{equation}
Y^2 = 1\,.
\end{equation}
{}From this we can derive
\begin{eqnarray}
(1-Y) E (1 + U) & = &  0\label{eq:yu}\\
(1+Y) E^{-1} (1 - U) & = &  0\,. \nonumber
\end{eqnarray}

We will find it useful to decompose all these matrices into two by two
block matrices associated with odd and even indices, as is done in
(\ref{eq:matX}).  We thus have
\begin{equation}
X =
\left(\begin{array}{cc}
0&  X_{oe}\\
X_{eo}& 0
\end{array} \right)\, ,
\end{equation}
where for example $X_{oe} = (X_{2k + 1, 2m})$.
Similarly, we have
\begin{equation}
C=
\left(\begin{array}{cc}
-1& 0 \\
0& 1
\end{array} \right),
\end{equation}
\begin{equation}
U=
\left(\begin{array}{cc}
U_{oo}& U_{oe}\\
U_{eo}& U_{ee}
\end{array} \right),\;\;\;\;\;
\bar{U}=
\left(\begin{array}{cc}
U_{oo}& -U_{oe}\\
-U_{eo}& U_{ee}
\end{array} \right),
\end{equation}
\begin{equation}
V=
\left(\begin{array}{cc}
V_{oo}&0 \\
0&V_{ee}
\end{array} \right)
= \frac{1}{3} \left(\begin{array}{cc}
-1 + 2 U_{oo}&0 \\
0&1 + 2 U_{ee}
\end{array} \right)\, ,
\label{eq:v2}
\end{equation}
\begin{equation}
Z=
\left(\begin{array}{cc}
-V_{oo}&0 \\
0&V_{ee}
\end{array} \right)\, .
\end{equation}

Using this block notation, we can use (\ref{eq:yu}) to derive some
further relations on $U_{oo}, U_{ee}$ which will be useful later.
{}From the $oo, oe$ components of the two relations in (\ref{eq:yu}) we
have
\begin{eqnarray}
(1 + U_{oo})-\sqrt{3}iMU_{eo} & = &  0\, ,\nonumber\\
-\sqrt{3}iM (1 + U_{ee}) + U_{oe} & = &  0\label{eq:u-relations}\, ,\\
3 (1-U_{oo}) -\sqrt{3} i (M^T)^{-1} U_{eo} & = &0\, ,\nonumber\\
\sqrt{3} i(M^T)^{-1} (1-U_{ee}) -3U_{oe} & = &0\, ,\nonumber
\end{eqnarray}
where we have defined
\begin{equation}
M = E^{-1} X_{oe} E, \;\;\;\;\;
(M^T)^{-1} = E X_{oe} E^{-1}\,.
\label{eq:m}
\end{equation}
{}From (\ref{eq:u-relations}) we have
\begin{eqnarray}
1 + U_{oo} & = &  3M (1 + U_{ee})M^T  \label{eq:U-relation}\\
1-U_{oo} & = & \frac{1}{3}  (M^T)^{-1} (1-U_{ee}) M^{-1}\,.\nonumber
\end{eqnarray}
{}From these we have
\begin{equation}
1-U_{oo}^2 = (M^T)^{-1} (1-U_{ee}^2) M^T\, ,
\end{equation}
which implies
\begin{equation}
\sqrt{1-U_{oo}^2} = (M^T)^{-1}  \sqrt{1-U_{ee}^2} M^T\,.
\label{eq:root-relation}
\end{equation}

We now wish to use these relations to demonstrate that $| \Xi_{-1} \rangle$
is a projection operator on the space of half-string functionals.
Since $T$ is a function of $Z$, and $Z_{nm}$ vanishes when $n + m$ is
odd, condition (a) from Section \ref{sec:rank1} is clearly satisfied.
It remains to demonstrate that condition (b) is satisfied.  Condition
(b) states that the matrix
\begin{equation}
L = E^{-1} \left( \frac{1-S}{ 1 + S}  \right) E^{-1}\, ,
  \label{eq:l-instanton}
\end{equation}
satisfies
\begin{equation}
L_{ee} = X_{eo} L_{oo} X_{oe}\,.
\label{eq:l-need}
\end{equation}
It is reasonable that such a condition should be satisfied, since as
we have seen $L$ is defined as a function of $X$.  Let us now verify
this explicitly.

{}From (\ref{eq:t}) we have
\begin{equation}
S= CT =
\left(\begin{array}{cc}
\frac{ 1}{2V_{oo}}
\left[ 1-V_{oo}-\sqrt{(1-3V_{oo}) (1 + V_{oo})}  \right]&0 \\
0&\frac{ 1}{2V_{ee}}
\left[ 1+V_{ee}-\sqrt{(1+3V_{ee}) (1 -V_{ee})}  \right]
\end{array} \right).
\end{equation}
This gives us
\begin{equation}
1 + S=
\left(\begin{array}{cc}
\frac{ 1}{2V_{oo}}
\sqrt{1 + V_{oo}}
\left[ \sqrt{1+V_{oo}}-\sqrt{1-3V_{oo}}\right] &0 \\
0&\frac{ 1}{2V_{ee}}
\sqrt{1 + 3V_{ee}}
\left[ \sqrt{ 1+ 3V_{ee}}-\sqrt{(1 -V_{ee})}  \right]
\end{array} \right)\, ,
\end{equation}
and
\begin{equation}
1 -S=
\left(\begin{array}{cc}
\frac{ 1}{2V_{oo}}
\sqrt{1 -3V_{oo}}
\left[ \sqrt{1+V_{oo}}-\sqrt{1-3V_{oo}}  \right]&0 \\
0&\frac{ 1}{2V_{ee}}
\sqrt{1 -V_{ee}}
\left[ \sqrt{ 1+ 3V_{ee}}-\sqrt{(1 -V_{ee})}  \right]
\end{array} \right)\, ,
\end{equation}
so that
\begin{equation}
L= E^{-1}
\left(\begin{array}{cc}
\frac{\sqrt{1 -3V_{oo}}}{\sqrt{1 +V_{oo}}} &0 \\
0&\frac{\sqrt{1 -V_{ee}}}{\sqrt{1 + 3V_{ee}}}
\end{array} \right)E^{-1}\,.
\end{equation}
Using (\ref{eq:v2}) this becomes
\begin{equation}
L= E^{-1}
\left(\begin{array}{cc}
\sqrt{3}
\frac{\sqrt{1 -U_{oo}}}{\sqrt{1 +U_{oo}}} &0 \\
0&\frac{1}{ \sqrt{3}} \frac{\sqrt{1 -U_{ee}}}{\sqrt{1 + U_{ee}}}
\end{array} \right) E^{-1}\,.
\label{eq:l-u}
\end{equation}

To show that $| \Xi_{-1} \rangle$ is a projector, we must prove
(\ref{eq:l-need}), which from (\ref{eq:l-u}) means that we must have
\begin{equation}
3 M^T \sqrt{\frac{1 -U_{oo}}{1 + U_{oo}}} M =
\sqrt{\frac{1 -U_{ee}}{1 + U_{ee}}}\, ,
\label{eq:xi-equation}
\end{equation}
where $M$ is defined in (\ref{eq:m}).  We can rearrange this equation
into the form
\begin{equation}
3 \frac{1}{ \sqrt{1-U_{ee}}}  M^T \sqrt{1-U_{oo}}
= \frac{1}{ \sqrt{1+U_{ee}}}  M^{-1} \sqrt{1+U_{oo}}\,.
\end{equation}
This leads to
\begin{equation}
3 \frac{1}{ \sqrt{1-U_{ee}^2}}  M^T \sqrt{1-U_{oo}^2}
= \frac{1}{ 1+U_{ee}}  M^{-1} (1+U_{oo}),
\end{equation}
which follows directly from (\ref{eq:U-relation}) and
(\ref{eq:root-relation}).

We have now shown that the state $| \Xi_{-1} \rangle$, corresponding
to a D-instanton in the RSZ model, satisfies the relations (a) and (b)
from Section \ref{sec:rank1} which are needed for a Gaussian
functional to act as a rank one projection on the space of half-string
functionals.  For the state $| \Xi_{-1} \rangle$ to be a normalizable
state in the Fock space, it is furthermore necessary that the
eigenvalues of $S$ be less than one and be square summable.  This in
turn requires that the spectrum of $Z$ lie in the range $(-1/3, 1)$
and approach zero.  This condition was checked numerically for the
sliver state in \cite{Kostelecky-Potting}.  We have verified
numerically that the spectrum of $Z$ for the D-instanton state $|
\Xi_{-1} \rangle$ is similar to that of the sliver state and seems to
be such that this state also is normalizable and in the Fock space
\footnote{Note added: It may be, however, that $S$ nonetheless has an
eigenvalue of -1, taking the D-instanton sliver as well as the
original sliver state slightly outside the Hilbert space
\cite{rsz-4,Sen-personal}}.  Thus, we have found that the state $|
\Xi_{-1} \rangle$ corresponds to a normalizable state in the string
Fock space which acts as a rank one projection on half-string
functionals.

\section*{Acknowledgements}

Thanks to M.\ Douglas, G.\ Moore, L.\ Rastelli, A.\ Sen, I.\ Singer,
and B.\ Zwiebach for helpful discussions and correspondence.  WT would
like to thank the ITP, Santa Barbara, and the ITP workshop on M-theory
for support and hospitality during the progress of this work.
The work of DJG was supported by the NSF under  the
grants  PHY 99-07949 and PHY 97-22022. The work of WT was supported in
part by the A.\ P.\ Sloan Foundation and in part by the DOE through
contract \#DE-FC02-94ER40818.

\normalsize

\bibliographystyle{plain}

\begin{thebibliography}{10}

\bibitem{Witten-SFT}
E.\ Witten, ``Non-commutative geometry and string field theory,'' \NP {\bf
   B268} (1986) 253.

\bibitem{Sen-universality}
A.\ Sen, ``Universality of the tachyon potential,'' {\it JHEP} {\bf 9912}
   (1999) 027, {\tt hep-th/9911116}.


\bibitem{level-truncation}
V.\ A.\ Kostelecky and S.\ Samuel,
\NP {\bf B336} (1990) 263-296;
A.\ Sen, B.\ Zwiebach
{\it
   JHEP} {\bf 0003} (2000) 002, {\tt hep-th/9912249};
N.~Moeller and W.~Taylor,
Nucl.\ Phys.\ B {\bf 583}, 105 (2000),
{\tt hep-th/0002237};
N.~Berkovits,
JHEP {\bf 0004}, 022 (2000),
{\tt hep-th/0001084};
J.~A.~Harvey and P.~Kraus,
JHEP {\bf 0004}, 012 (2000),
{\tt hep-th/0002117};
N.~Berkovits, A.~Sen and B.~Zwiebach,
Nucl.\ Phys.\ B {\bf 587}, 147 (2000),
{\tt hep-th/0002211};
R.~de Mello Koch, A.~Jevicki, M.~Mihailescu and R.~Tatar,
Phys.\ Lett.\ B {\bf 482}, 249 (2000),
{\tt hep-th/0003031};
P.~De Smet and J.~Raeymaekers,
JHEP {\bf 0005}, 051 (2000),
{\tt hep-th/0003220};
A.~Iqbal and A.~Naqvi,
{\tt hep-th/0004015};
P.~De Smet and J.~Raeymaekers,
JHEP {\bf 0008}, 020 (2000),
{\tt hep-th/0004112};
N.~Moeller, A.~Sen and B.~Zwiebach,
JHEP {\bf 0008}, 039 (2000),
{\tt hep-th/0005036};
J.~R.~David,
JHEP {\bf 0010}, 017 (2000),
{\tt hep-th/0005085};
A.~Sen and B.~Zwiebach,
JHEP {\bf 0010}, 009 (2000),
{\tt hep-th/0007153};
J.~R.~David,
JHEP {\bf 0010}, 004 (2000),
{\tt hep-th/0007235};
R.~de Mello Koch and J.~P.~Rodrigues,
Phys.\ Lett.\ B {\bf 495}, 237 (2000),
{\tt hep-th/0008053};
W.~Taylor,
JHEP {\bf 0008}, 038 (2000),
{\tt hep-th/0008033};
N.~Moeller,
{\tt hep-th/0008101};
A.~Iqbal and A.~Naqvi,
JHEP {\bf 0101}, 040 (2001),
{\tt hep-th/0008127};
H.~Hata and S.~Shinohara,
JHEP {\bf 0009}, 035 (2000),
{\tt hep-th/0009105};
P.~Mukhopadhyay and A.~Sen,
JHEP {\bf 0102}, 017 (2001),
{\tt hep-th/0101014};
H.~Hata and S.~Teraguchi,
{\tt hep-th/0101162};
I.~Ellwood and W.~Taylor,
{\tt hep-th/0103085};
B.~Feng, Y.~He and N.~Moeller,
{\tt hep-th/0103103};
I.~Ellwood, B.~Feng, Y.~He and N.~Moeller,
{\tt hep-th/0105024}.



\bibitem{large-B}
K.~Dasgupta, S.~Mukhi and G.~Rajesh,
JHEP {\bf 0006}, 022 (2000),
{\tt [hep-th/0005006};
J.~A.~Harvey, P.~Kraus, F.~Larsen and E.~J.~Martinec,
JHEP {\bf 0007}, 042 (2000),
{\tt hep-th/0005031};
E.~Witten,
{\tt hep-th/0006071};
R.~Gopakumar, S.~Minwalla and A.~Strominger,
{\tt hep-th/0007226};
N.~Seiberg,
JHEP {\bf 0009}, 003 (2000),
{\tt hep-th/0008013};
J.~A.~Harvey and G.~Moore,
{\tt hep-th/0009030};
A.~Sen,
JHEP {\bf 0011}, 035 (2000),
{\tt hep-th/0009038};
R.~Tatar,
{\tt hep-th/0009213};
P.~Kraus, A.~Rajaraman and S.~Shenker,
Nucl.\ Phys.\ B {\bf 598}, 169 (2001),
{\tt hep-th/0010016};
L.~Cornalba,
Phys.\ Lett.\ B {\bf 504}, 55 (2001),
{\tt hep-th/0010021};
K.~Okuyama,
Phys.\ Lett.\ B {\bf 499}, 167 (2001),
{\tt hep-th/0010028};
M.~Schnabl,
JHEP {\bf 0011}, 031 (2000),
{\tt hep-th/0010034};
J.~A.~Harvey, P.~Kraus and F.~Larsen,
JHEP {\bf 0012}, 024 (2000),
{\tt hep-th/0010060};
O.~Andreev,
Nucl.\ Phys.\ B {\bf 598}, 151 (2001),
{\tt hep-th/0010218};
T.~Krajewski and M.~Schnabl,
{\tt hep-th/0104090}.


\bibitem{bi}
A.~A.~Gerasimov and S.~L.~Shatashvili,
JHEP {\bf 0010}, 034 (2000),
{\tt hep-th/0009103};
D.~Kutasov, M.~Marino and G.~Moore,
JHEP {\bf 0010}, 045 (2000),
{\tt hep-th/0009148};
D.~Ghoshal and A.~Sen,
JHEP {\bf 0011}, 021 (2000),
{\tt hep-th/0009191};
D.~Kutasov, M.~Marino and G.~Moore,
{\tt hep-th/0010108};
S.~Moriyama and S.~Nakamura,
Phys.\ Lett.\ B {\bf 506}, 161 (2001),
{\tt hep-th/0011002};
A.~A.~Gerasimov and S.~L.~Shatashvili,
JHEP {\bf 0101}, 019 (2001),
{\tt hep-th/0011009};
I.~Y.~Aref'eva, A.~S.~Koshelev, D.~M.~Belov and P.~B.~Medvedev,
{\tt hep-th/0011117};
P.~Kraus and F.~Larsen,
Phys.\ Rev.\ D {\bf 63}, 106004 (2001),
{\tt hep-th/0012198};
M.~Alishahiha, H.~Ita and Y.~Oz,
Phys.\ Lett.\ B {\bf 503}, 181 (2001),
{\tt hep-th/0012222};
G.~Chalmers,
{\tt hep-th/0103056};
M.~Marino,
{\tt hep-th/0103089};
V.~Niarchos and N.~Prezas,
{\tt hep-th/0103102};
K.~S.~Viswanathan and Y.~Yang,
{\tt hep-th/0104099};
M.~Alishahiha,
{\tt hep-th/0104164}.


\bibitem{other}
K.\ Bardakci and M.\ B.\ Halpern,
\PR {\bf  D10} (1974) 4230;
K.\ Bardakci,
\NP {\bf B133} (1978) 297;
O.~Bergman, K.~Hori and P.~Yi,
Nucl.\ Phys.\ B {\bf 580}, 289 (2000),
{\tt hep-th/0002223};
J.~A.~Harvey, D.~Kutasov and E.~J.~Martinec,
{\tt hep-th/0003101};
D.~Ghoshal and A.~Sen,
Nucl.\ Phys.\ B {\bf 584}, 300 (2000),
{\tt hep-th/0003278};
T.~Suyama,
Phys.\ Lett.\ B {\bf 488}, 83 (2000),
{\tt hep-th/0006052};
B.~Zwiebach,
JHEP {\bf 0009}, 028 (2000),
{\tt hep-th/0008227};
J.~A.~Minahan and B.~Zwiebach,
JHEP {\bf 0009}, 029 (2000),
{\tt hep-th/0008231};
T.~Suyama,
{\tt hep-th/0008242};
G.~Gibbons, K.~Hori and P.~Yi,
Nucl.\ Phys.\ B {\bf 596}, 136 (2001),
{\tt hep-th/0009061};
A.~Sen,
JHEP {\bf 0012}, 001 (2000),
{\tt hep-th/0009090};
J.~Kluson,
JHEP {\bf 0011}, 016 (2000),
{\tt hep-th/0009189};
K.~Bardakci and A.~Konechny,
Nucl.\ Phys.\ B {\bf 598}, 427 (2001),
{\tt hep-th/0009214};
J.~A.~Minahan and B.~Zwiebach,
JHEP {\bf 0103}, 038 (2001),
{\tt hep-th/0009246};
B.~Zwiebach,
{\tt hep-th/0010190};
A.~Sen,
{\tt hep-th/0010240};
S.~Dasgupta and T.~Dasgupta,
{\tt hep-th/0010247};
A.~A.~Tseytlin,
{\tt hep-th/0011033};
G.~Mandal and S.~R.~Wadia,
Nucl.\ Phys.\ B {\bf 599}, 137 (2001),
{\tt hep-th/0011094};
J.~A.~Minahan and B.~Zwiebach,
JHEP {\bf 0102}, 034 (2001),
{\tt hep-th/0011226};
M.~Schnabl,
Phys.\ Lett.\ B {\bf 504}, 61 (2001),
{\tt hep-th/0011238};
G.~Arutyunov, S.~Frolov, S.~Theisen and A.~A.~Tseytlin,
JHEP {\bf 0102}, 002 (2001),
{\tt hep-th/0012080};
M.~Kleban, A.~E.~Lawrence and S.~Shenker,
{\tt hep-th/0012081};
J.~A.~Minahan,
JHEP {\bf 0103}, 028 (2001),
{\tt hep-th/0102071};
K.~Hashimoto and S.~Hirano,
JHEP {\bf 0104}, 003 (2001),
{\tt hep-th/0102173};
K.~Hashimoto and S.~Hirano,
{\tt hep-th/0102174}.




\bibitem{Gross-Taylor-II}
D.\ J.\ Gross and W.\ Taylor,
``Split string field theory  II,''
{\tt hep-th/0106036}


\bibitem{rsz}
L.~Rastelli, A.~Sen and B.~Zwiebach,
``String field theory around the tachyon vacuum,'',
{\tt hep-th/0012251}.

\bibitem{rsz-2}
L.~Rastelli, A.~Sen and B.~Zwiebach,
``Classical solutions in string field theory around the tachyon vacuum,'',
{\tt hep-th/0102112}.

\bibitem{Kostelecky-Potting}
V.~A.~Kostelecky and R.~Potting,
``Analytical construction of a nonperturbative vacuum for the open 
bosonic string,''
Phys.\ Rev.\ D {\bf 63}, 046007 (2001),
{\tt hep-th/0008252}.


\bibitem{Rastelli-Zwiebach}
L.~Rastelli and B.~Zwiebach,
``Tachyon potentials, star products and universality,''
{\tt hep-th/0006240}.



\bibitem{rsz-3}
L.~Rastelli, A.~Sen and B.~Zwiebach,
``Half-strings, projectors, and multiple D-branes in vacuum string field
theory,''
{\tt hep-th/0105058}.

\bibitem{Gross-Jevicki-12}
D.\ J.\ Gross and A.\ Jevicki, ``Operator formulation of interacting string
   field theory (I), (II),'' \NP {\bf B283} (1987) 1; \NP {\bf B287} (1987) 225.

\bibitem{bcnt}
J.\ Bordes, Chan H.-M., L.\ Nellen, Tsou S.-T.
``Half-string oscillator approach to string field theory,''
\NP {\bf
   B351} (1991)  441.

\bibitem{Abdurrahman-Bordes}
A.\ Abdurrahman and J.\ Bordes,
``The relationship between the comma theory and Witten's string field
theory,'' \PR {\bf D 58}:086003  (1998).

\bibitem{Harvey-lectures}
J.~A.~Harvey,
``Komaba lectures on noncommutative solitons and D-branes,''
hep-th/0102076.

\bibitem{rsz-4}
L.~Rastelli, A.~Sen and B.~Zwiebach,
``Boundary CFT construction of D-branes in vacuum string field theory,''
{\tt hep-th/0105168}.


\bibitem{Sen-personal}
L.~Rastelli, A.~Sen and B.~Zwiebach,
personal communication.



\end{thebibliography}

\end{document}